\documentclass[twocolumn,showpacs,preprintnumbers,amsmath,amssymb]{revtex4}

\usepackage{graphicx}
\usepackage{dcolumn}
\usepackage{bm}

\begin{document}


\title{Higgsless superconductivity from topological defects in compact BF terms}

\author{M. Cristina Diamantini}
\email{cristina.diamantini@pg.infn.it}
\affiliation{%
INFN and Dipartimento di Fisica, University of Perugia, via A. Pascoli, I-06100 Perugia, Italy
}%


\author{Carlo A. Trugenberger}
\email{ca.trugenberger@bluewin.ch}
\affiliation{%
SwissScientific, chemin Diodati 10, CH-1223 Cologny, Switzerland
}%


\date{\today}

\begin{abstract}
We present a new Higgsless model of superconductivity, inspired from anyon superconductivity but P- and T-invariant and generalizable to any dimension. While the original anyon superconductivity mechanism was based on incompressible quantum Hall fluids as average field states, our mechanism involves topological insulators as average field states. In D space dimensions it involves a (D-1)-form fictitious pseudovector gauge field which originates from the condensation of topological defects in compact low-energy effective BF theories. In the average field approximation, the corresponding uniform emergent charge creates a gap for the (D-2)-dimensional branes via the Magnus force, the dual of the Lorentz force. One particular combination of intrinsic and emergent charge fluctuations that leaves the total charge distribution invariant constitutes an isolated gapless mode leading to superfluidity. The remaining massive modes organise themselves into a D-dimensional charged, massive vector. There is no massive Higgs scalar as there is no local order parameter. When electromagnetism is switched on, the photon acquires mass by the topological BF mechanism. Although the charge of the gapless mode (2) and the topological order (4) are the same as those of the standard Higgs model, the two models of superconductivity are clearly different since the origins of the gap, reflected in the high-energy sectors are totally different. In 2D this type of superconductivity is explicitly realized as global superconductivity in Josephson junction arrays. In 3D this model predicts a possible phase transition from topological insulators to Higgsless superconductors.
\end{abstract}
\maketitle

\section{Introduction}
Anyon superconductivity \cite{wil} was briefly very popular in the early 90s as a possible mechanism for the high-$T_c$ cuprates. The idea, based on the possibility of fractional statistics \cite{wil} in 2 space dimensions (2D), involves fermions interacting with a fictitious, statistical Chern-Simons (CS) gauge field \cite{jac} which turns them into anyons by attaching magnetic flux to their charge density. In the average field approximation the fermions feel a collective, uniform fictitious magnetic field. If the Chern-Simons coupling constant is chosen so that an integer number of Landau levels are filled, the fermion state has a gap. A specific coherent fluctuation of the fermion density {\it together} with the statistical gauge field, however, is gapless. This massless mode is protected from decaying into particle-hole excitations by the average field gap and leads to anyon superfluidity. It can be shown that the origin of the massless mode is not the spontaneous breaking of a symmetry but, rather, the necessary restoration of the commutativity of translations which is broken in the average field approximation \cite{gid}. The most studied case is that of two filled Landau levels, corresponding to semions, or half fermions since, in this case, the charge order parameter of the resulting superconductor is 2 \cite{wil}. Unfortunately, the high-$T_c$ cuprates do not exhibit the P and T violation necessarily implied by the superconductivity mechanism based on anyons and thus the idea of anyon superconductivity was quickly abandoned. 

In this paper we revisit the anyon superconductivity mechanism to show that it can be made P- and T-invariant and extended to three space dimensions (3D), where it can be realized upon a phase transition from topological insulators \cite{topins}. To this end we start from the pure gauge formulation of \cite{zee}. We do not consider the usual case of semions (half-fermions) but, rather, we concentrate on the simple case in which the statistical interaction turns the fermions into bosons. 
By soldering together two fermion fluids of opposite spin, interacting with the same fictitious statistical gauge field and filling their respective first Landau levels in the average field approximation, one obtains a single gapless mode with charge 2. The remaining degrees of freedom organize themselves into a massive {\it vector} particle (with two polarisations in 2D) of unit charge, rather than a single neutral, scalar Higgs boson. This massive vector is described by two Chern-Simons terms of opposite parity and the same absolute value of the coupling constant. The P and T symmetries are preserved. 

We show that this type of P- and T-invariant Higgsless superconductivity can be reformulated by a rotation of the degrees of freedom as a charge and a vortex fluids interacting with each other by a mutual Chern-Simons term. The vortices interact additionally with a pseudovector statistical gauge field which, however, has no self-action. In this formulation, superfluidity arises as follows. In the average field approximation the statistical gauge field provides a uniform charge for the vortices. A charge arises instead of a magnetic field as in standard anyon superconductivity since here the statistical gauge field is a pseudovector. Note, however, that this emergent charge has to be distinguished from the intrinsic charge coupling to real electromagnetic fields, as has been stressed in \cite{moe}. The vortices are subject to the Magnus force \cite{magnus}, which, in 2D, is the exact dual of the Lorentz force and they are thus quantized into (dual) Landau levels. When an integer number of Landau levels is filled the vortices have a gap; this happens in particular for the ground state configuration with no vortices at all. This gap for vortices is nothing else than the Meissner effect. Due to the mutual Chern-Simons term, a generic intrinsic charge fluctuation corresponds to a local variation in the charge distribution felt by the vortices. Exactly as in the original mechanism of anyon superconductivity, this must be accompanied by an excitation of vortex-antivortex pairs through the Landau level gap \cite{chen}. Generic intrinsic charge fluctuation are thus also gapped. If, however the intrinsic charge fluctuation is accompanied by an emergent charge fluctuation such that the total charge felt by vortices remains unchanged, there is no energy price to pay. This is  the superfluid gapless mode. 

When electromagnetism is switched on, the photon acquires mass through the topological Chern-Simons mass mechanism \cite{jac}. The resulting completely gapped state has topological order \cite{wen} characterised by a ground state degeneracy $4^g$ on Riemann surfaces of genus $g$.

The important point is, finally that, in the mutual Chern-Simons formulation, this Higgsless topological superconductivity mechanism is easy to generalize to any number of dimensions \cite{dst2} and, in particular to three space dimensions (3D).  Indeed, a mutual Chern-Simons term can be generalized to any number of dimensions as a topological BF term \cite{birmi}. In 3D, the BF term involves the topological coupling of a standard vector gauge field to a two-form Kalb-Ramond gauge field \cite{kalb} with generalized vector gauge symmetry. Vortices are themselves described by a conserved antisymmetric tensor current and the emergent gauge field coupling to them is also a two-form gauge field. The dual "magnetic field" associated to a pseudotensor gauge field $\alpha_{\mu \nu}$ is the scalar density $F^0$, where $F^{\mu } = \epsilon^{\mu \nu \beta \sigma} \partial_{\nu}\alpha_{\beta \sigma}$. In the average field approximation, this corresponds again to a uniform emergent charge $F^0$ for the vortices. We show that such a uniform Kalb-Ramond scalar density leads to a gap for vortex strings via the Magnus force: this causes the Meissner effect in 3D. The superfluidity mechanism in 3D parallels thus verbatim the 2D situation: vortex strings are gapped in the average field approximation; generic intrinsic charge fluctuations are also gapped since they modify locally the uniform charge distribution felt by the vortices and cause thus the nucleation of gapped vortex strings; a specific combination of intrinsic and emergent charge fluctuations that leaves the scalar Kalb-Ramond density invariant, however, is the gapless superfluid mode. The remaining degrees of freedom represent a massive vector (spin 1) boson via the BF topological mass mechanism \cite{bow}; also in this case there is no scalar Higgs boson. Of course, what does not carry over obviously from 2D, instead, is the original interpretation of the emergent gauge field as a statistics changing interaction. 

The BF term represents the low-energy physics of topological insulators \cite{moore} (both in 2D and 3D). The relation to the original mechanism of anyon superconductivity is thus particularly suggestive. There, the gapped average field state was an incompressible quantum Hall fluid, here it is a topological insulator. Originally, the additional gauge field leading to the gapless superfluid mode was the statistics changing Chern-Simons gauge field, attaching fluxes to charges. Here the additional gauge field is a pseudovector and it attaches emergent charges to intrinsic charges, thereby preserving the P and T symmetries. While such a gauge field can be introduced in any number of dimensions, it cannot arise from fractional statistics in higher dimensions. 

The origin of the emergent gauge field lies, rather in the compactness of the BF term \cite{dst2}. In 2D, the BF term reduces to the mutual Chern-Simons interaction and this represents \cite{dst1} the physics of Josephson junction arrays \cite{faz}. When the two U(1) gauge groups are compact, however, the BF theory has to be formulated as a cutoff theory, e.g. on the lattice, with the unavoidable presence of topological defects. As is well known \cite{kog}, naively irrelevant operators like the Maxwell terms for the two emergent gauge fields can lead to non-perturbative effects if the corresponding masses lie below the cutoff. This is exactly what happens in Josephson junction arrays: the condensation of "electric" topological defects provides the statistical gauge field and global superconductivity in the array is exactly an example of the Higgsless topological superconductivity described above. 

As already pointed out, the BF term represents the low-energy physics of topological insulators \cite{moore}. As an effective low-energy theory of a condensed matter system, it must necessarily be considered as a cutoff theory with compact gauge groups. This implies additional topological excitations in the action. When these are dilute, the resulting physics is indeed that of topological insulators, in which both charges and vortices are screened in the bulk by the topological BF mass $m_{\rm BF}$. When the topological excitations condense, however, there is a phase transition to the Higgsless topological superconductivity described here. In 3D, the Maxwell term for the usual vector gauge field is {\it marginal} and must be added anyhow to the action to establish the complete phase diagram \cite {dst3}. The dynamical term for the Kalb-Ramond gauge field, instead involves an antisymmetric three-tensor. When these terms are included one can easily prove by standard free energy arguments that a phase transition from a topological insulator to a Higgless superconductor takes indeed place when the charge screening in the topological insulator becomes strong enough, $m_{\rm BF}/\Lambda > \lambda_{\rm crit}$, where $\Lambda$ is the ultraviolet cutoff and $\lambda_{\rm crit} \ll 1$. It is not clear to us at this point how the Higgsless superconductors introduced here are related to the "topological superconductors" described e.g. in \cite{topsup}. This is under current investigation. 

We would like to stress that the superconductivity model presented here is genuinely different than both the BCS dynamical symmetry breaking by pairing and its Abelian Higgs model description. In the present case the components of a "pair" in 2D are already bosons by themselves, having been individually transmuted from fermions by the statistical gauge interaction. It is not the paired fermions that produce a charge condensate via Bose-Einstein condensation, the "condensate" is provided by the emergent charge of the pseudovector statistical gauge field. Neither can our model be phenomenologically described by the Abelian Higgs model, although they share the same charge 2 and the same topological order 4 \cite{han}: in the present model there is no local order parameter and, correspondingly, there is no scalar Higgs boson, but rather a massive charged vector. The properties of the present model are, rather, suggestive of a possible connection with Higgsless models of spontaneous symmetry breaking via extra dimensions and the AdS/CFT correspondence \cite{csa}. 

Also note that this model is an example of the fact that one cannot deduce the characteristics of a superconductivity model by looking at the low-energy (photon plus gapless mode) sector exclusively. 
Doing so in this case would have lead to the erroneous conclusion that this model coincides with the abelian Higgs model, since both the charge of the gapless mode and the topological order are the same. Instead, in order to establish the nature of a superconductivity model it is crucial to focus on the mechanism by which the gap is opened and the gapless mode is originated and this is intimately connected with the high-energy sector. In a sense, all possible superconductivity models must have the same phenomenological low-energy structure, it is only the origin of the gap, and thus the high-energy physics, that can distinguish them. 

This paper is organised as follows. In section 2 we revisit standard anyon superconductivity. In sections 3 and 4 we introduce the doubled anyon superconductivity model and its mutual Chern-Simons version. In section 5 we generalize the model to any number of dimensions and we derive the origin of the emergent gauge field as a topological excitation in compact BF models. In section 6 we focus on 3D and the transition from topological insulators to Higgsless superconductors. Finally we draw our conclusions in section 7.

\section{Anyon superconductivity revisited}
Anyons are particles of fractional statistics in two space dimensions (2D) \cite{wil}. As is well known, the deviation from standard boson or fermion statistics can be described by the interaction with a fictitious statistical gauge field $\alpha_{\mu}$ whose action is given by the topological Chern-Simons term \cite{jac}.  Consider for example the following (non relativistic) Lagrangian density (we shall use units in which $c=1$, $\hbar=1$ and $e=1$ throughout the paper),
\begin{equation}
{\cal L} = \psi^{\dag} i\left( \partial_0 +i{\alpha}_0 \right) \psi +{1\over 2m}\psi^{\dag} \left( \partial_i +i{\alpha}_i\right)^2 \psi +{k\over 4\pi}
\alpha_{\mu}\epsilon^{\mu \nu \sigma} \partial_{\nu}\alpha_{\sigma} \ .
\label{aone}
\end{equation}
The equation of motions for $\alpha_0$ requires
\begin{equation}
\rho \equiv \psi^{\dag} \psi = {k\over 2\pi} \epsilon^{ij}\partial_i \alpha_j =  {k\over 2\pi} \cal{B} \ ,
\label{atwo}
\end{equation}
where $\cal{B}$ is the fictitious statistical magnetic field and the sign of the Chern-Simons coupling constant $k$ is the same as that of $\cal{B}$. For simplicity we shall assume that ${\cal{B}}> 0$ so that also $k > 0$. Each particle carries thus $1/k$ units of statistical magnetic field. This attachment contributes a phase $2\pi/ k$ to the wave function of a particle when it is carried around another one, or to a statistics phase $\pi / k$ when two particles are interchanged. If the original particles are fermions, as we shall assume henceforth, the total statistics phase is $\theta = \pi (1-1/k)$. 

Let us now consider the average field approximation, in which the statistical magnetic field of a uniform particle distribution is substituted by its uniform average. Suppose that the particle density is $\rho$; the particles will then fall into Landau levels with a filling fraction $\nu = 2\pi \rho /\cal{B}$. But the statistical magnetic field is itself tied to the particle density by the Chern-Simons equation of motion (\ref{atwo}) and thus $\nu = k$. When $k = m \in \mathbb{N}$ exactly $m$ Landau levels are filled and the average field state is gapped. In particular, if we add a real magnetic field $B$ this will disturb the self-consistent Landau level balance and some particles or holes will be excited across the Landau level gap \cite{chen}: this gap for real magnetic fields is essentially the Meissner effect. 

Before proceeding beyond the average field approximation let us pause to ask ourselves if there are also other gapped states with uniform density. This is indeed the case: all states of the Jain hierarchy \cite{jain} of fractional quantum Hall states with filling fraction $\nu = m/(mp+1)$ with $p$ an even integer are also gapped. The low-energy degrees of freedom for these fractional quantum Hall states can be described in terms of a Chern-Simons field theory for $m$ pseudovector gauge fields $a^i_{\mu}$, $i=1 \dots m$,  grouped into an "isospin" vector ${\bf a}_{\mu} = \left( a^1_{\mu} \dots a^m_{\mu} \right)$ such that 
\begin{equation}
j^{\mu} = \sum_{i=1}^m \left( j^{\mu} \right)^i= (1/2\pi )\sum_{i=1}^m \ \epsilon^{\mu \nu \sigma} \partial_{\nu}  a^i_{\sigma} \
\label{athree}
\end{equation}
represents the conserved current of electrons \cite{wen}. The Lagrangian density for this Chern-Simons theory can be represented compactly as 
\begin{equation}
{\cal L} = {1\over 4\pi} {\bf a}^T_{\mu} K \epsilon^{\mu \nu \sigma} \partial_{\nu}{\bf a}_{\sigma} + j^{\mu} A_{\mu} \ ,
\label{afour}
\end{equation}
where we have included the coupling to the real electromagnetic gauge potential $A_{\mu }$ and the matrix $K$ is given by
\begin{equation}
K = 
\begin{pmatrix}
  p+1 & p & \cdots & p \\
  p & p+1 & \cdots & p\\
  \vdots  & \vdots  & \ddots & \vdots  \\
  p & p & \cdots & p+1
\end{pmatrix}
\ .
\label{afive}
\end{equation}
The inclusion of the electromagnetic coupling completely fixes the overall normalization of the matter gauge fields $a^i_{\mu}$. The idea is that (\ref{afour}) represents the infrared-dominant term in the low-energy effective action for the electrons written in terms of effective gauge fields. The next terms in the effective gauge theory for the electrons would be Maxwell terms (possibly non-relativistic) but these are irrelevant operators in 2D. The Chern-Simons action (\ref{afour}) encodes the topological masses of matter fluctuations around the zeroth-order average-field approximation: since all eigenvalues of $K$ are different from zero, the average-field state is completely gapped. As has been pointed out in \cite{zee}, all the fractional quantum Hall fluids (\ref{afive}) can serve as average-field approximation for an anyon superconductor. 

Let us now move beyond the average field approximation and permit the fictitious statistical field to fluctuate around this zeroth-order configuration. To this end we have to couple back the statistical gauge field $\alpha_{\mu}$ as in (\ref{aone}), thereby changing $k$ into $m/(mp+1)$ to reflect the new filling fraction. We can simply introduce the $(m+1)$-dimensional vectors of ${\cal A}_{\mu} = \left( \alpha _{\mu }, a^1_{\mu} \dots a^m_{\mu} \right)$ and ${\bf A} = \left( 0, A_{\mu} \dots A_{\mu} \right)$, so that the fully coupled 
Lagrangian density is
\begin{equation}
{\cal L} = {1\over 4\pi} {\cal A}^T_{\mu} \Lambda \ \epsilon^{\mu \nu \sigma} \partial_{\nu}{\cal A}_{\sigma} + 
{1\over 2\pi} {\cal A}^T_{\mu} \epsilon^{\mu \nu \sigma} \partial_{\nu}{\bf A}_{\sigma} \ ,
\label{asix}
\end{equation}
with the new $m+1$ dimensional matrix $\Lambda$ given by
\begin{equation}
\Lambda = 
\begin{pmatrix}
  {m\over mp+1} & -1 & \cdots & \cdots & -1 \\
  -1 & p+1 & p & \cdots & p\\
  -1 & p & p+1 & \cdots & p \\
  \vdots  & \vdots  & \vdots & \ddots & \vdots  \\
  -1 & p & \cdots & \cdots & p+1
\end{pmatrix}
\ .
\label{aseven}
\end{equation}

A very interesting thing happens when the statistical gauge field is permitted to fluctuate around the zeroth-order average field approximation as in (\ref{asix}). Indeed, the new matrix $\Lambda$ has now exactly one vanishing eigenvalue corresponding to the charge $m$ eigenvector 
\begin{eqnarray}
\left( {\phi \over 2\pi}, q^i, \dots , q^m \right) &&= (mp+1,1,\dots, 1) \ ,
\nonumber \\
\phi &&\equiv \int d^2 x \ \cal{B}
\nonumber \\
q^i &&\equiv \int d^2 x \ (j^0)^i
\label{aeight}
\end{eqnarray}
This gapless mode is protected from mixing with a continuum of particle-hole excitations by the gap of the average-field approximation and represents the anyon superfluid mode that gives mass to the photon via the mixed ${\cal A}{\bf A}$ Chern-Simons term in (\ref{asix}). 

The focus of the interest in anyon superconductivity has been mostly on the case $m=2$, in which the particles are semions (half-fermions) of statistics $\theta = \pi/2$, since this would corresponds to a traditional order parameter 2. The Chern-Simons term in (\ref{asix}), however, breaks the discrete symmetries of parity (P ) under which the sign of one space component, say the first, of all three-vectors is reversed and the of time-reversal (T), under which the sign of the time component of all three-vectors is reversed,
\begin{eqnarray}
P &&: \ \ V^1 \to -V^1 \quad \forall \ \ {\rm three-vectors} \ V^{\mu} \ ,
\nonumber \\
T &&: \ \ V^0 \to -V^0 \quad \forall \ \ {\rm three-vectors} \ V^{\mu} \ .
\label{anine}
\end{eqnarray}
Note that, actually, the effective matter gauge fields in the first term of (\ref{asix}) are pseudovectors, for which the P and T symmetries require an additional minus sign, so that P changes the sign of the time component and the second space components of three vectors and T changes the sign of all space components of three vectors. This ensures that the electromagnetic coupling (the second term in (\ref{asix})) of an effective pseudovector gauge field to the real vector gauge respects both the P and T symmetries. The matter action (the first term in (\ref{asix})), however, breaks both discrete symmetries. Anyon superconductivity was quickly abandoned when experiment on the high-$T_c$ curates failed to detect any P and/or T violation.

\section{Doubled anyon superconductivity}
In this section we shall show how the original anyon superconductivity mechanism described above can be modified in a P- and T-invariant fashion easily generalizable to any number of dimensions and, in particular to 3D. 
Let us start from the anyon superfluid action (\ref{asix}) and (\ref{aseven}) for the simplest case $m=1$ of bosons. Even if we shall henceforth focus only on the fermion to boson transmutation we will continue speaking of "anyon superconductivity", meaning thereby the basic mechanism leading to superconductivity, which, as we will see, is distinct from the standard pairing mechanism
and its Higgs model description. 

The key observation is that, in the previous section, the spin 1/2 of the original fermions has been completely neglected. Now,
irreducible two-component spinors in 2D carry only half of the spin degree of freedom \cite{boy}, or, in other words, they have their spin pointing only in one direction, up or down. This means that irreducible 2D electrons carry a pseudoscalar (vortex) degree of freedom $S_z=\pm 1/2$, in addition to charge. Spin up electrons can thus be described by a gauge field $a^{+}$ in terms of which the current 
\begin{equation}
(\phi^{\mu})^{+} = {1\over 2\pi} \epsilon^{\mu \nu \sigma} \partial_{\nu} a_{\sigma}^{+} \ ,
\label{bone}
\end{equation}
carries both vorticity and charge, $\phi^{+}=\int d^2 x \ (\phi^0)^{+} = 2S_z^{+} + q^{+}$. We will now achieve the fermion to boson transmutation by coupling this current to a pseudovector (rather than a vector as usual) statistical gauge field. From now on we shall thus consider $\alpha_{\mu}$ as a pseudovector gauge field. This has the consequence that $(1/2\pi) \epsilon^{\mu \nu \sigma} \partial_{\nu}\alpha_{\sigma}$ represents a background charge, rather than magnetic flux. Note however, that this emergent statistical charge is different from the intrinsic charge that couples to physical electromagnetic fields, as has been stressed in \cite{moe}. 

In this representation, the fermion to boson transmutation is a consequences of carrying an emergent charge around a vortex represented by (twice) the electron spin, rather then the other way around, as in standard anyon superconductivity. Correspondingly, the quantization of a particle carrying vorticity in the background of a uniform statistical charge is completely equivalent to the usual case of the Landau levels of a particle carrying charge in a uniform magnetic field. This is because, in 2D, the Magnus force on vortices is the exact dual of the Lorentz force on charges \cite{magnus} (we will present below the detailed derivation of the Magnus force for the 3D case). The gap depends only on the quantity $|eB|$: in the usual case $e$ is a scalar and $B$ a pseudoscalar, in the present case it is the other way around, but the overall properties of the combination $eB$ remain the same. In the present case, however, particles carry both vorticity and intrinsic charge. The average field state is thus a (dual) incompressible fluid of particles carrying both vorticity and intrinsic charge in a uniform statistical charge distribution, analogous to the incompressible fluid of anyons in the Haldane-Halperin hierarchy \cite{hh}. The simplest realization of anyon superconductivity starting from electrons with spin up can thus be formulated as 
\begin{equation}
{\cal L}^{+} = {1\over 4\pi} ({\cal A}^{+})^T_{\mu} \Lambda^{+} \ \epsilon^{\mu \nu \sigma} \partial_{\nu}{\cal A}^{+}_{\sigma} + 
{1\over 2\pi} ({\cal A}^{+})^T_{\mu} \epsilon^{\mu \nu \sigma} \partial_{\nu}{\bf A}_{\sigma} \ ,
\label{btwo}
\end{equation}
with ${\cal A}^{+}_{\mu} = \left( \alpha _{\mu }, a^{+}_{\mu} \right)$ and 
\begin{equation}
\Lambda^{+} = 
\begin{pmatrix}
  {1\over p+1} & -1 \\
  -1 & p+1 \\
 \end{pmatrix}
\ .
\label{bthree}
\end{equation}
The matrix $\Lambda^{+}$ has the eigenvalues 0, corresponding to the superfluid mode and $\lambda^{+}=((p+1)^2 +1)/(p+1)$ describing the remaining massive mode. 

For $k=1/(p+1)$, with p even, the statistics parameters $\theta = \pi (1-1/k)= -p\pi $ and $\theta = \pi (1+1/k)=(2+p) \pi $ are equivalent since they differ by an integer multiple of $2\pi$. Therefore, one can also consider the parity-reversed model corresponding to 2D electrons with spin down. This is formulated in terms of the gauge field $a^{-}$ so that the current 
\begin{equation}
(\phi^{\mu})^{-} = {1\over 2\pi} \epsilon^{\mu \nu \sigma} \partial_{\nu} a_{\sigma}^{-} \ ,
\label{bbthree}
\end{equation}
carries combined vorticity and intrinsic charge, $\phi^{-}=  2S_z^{-} - q^{-}$, 
\begin{equation}
{\cal L}^{-} = {1\over 4\pi} ({\cal A}^{-})^T_{\mu} \Lambda^{-} \ \epsilon^{\mu \nu \sigma} \partial_{\nu}{\cal A}^{-}_{\sigma} - 
{1\over 2\pi} ({\cal A}^{-})^T_{\mu} \epsilon^{\mu \nu \sigma} \partial_{\nu}{\bf A}_{\sigma} \ ,
\label{bfour}
\end{equation}
with ${\cal A}^{-}_{\mu} = \left( \alpha _{\mu }, a^{-}_{\mu} \right)$ and 
\begin{equation}
\Lambda^{-} = 
\begin{pmatrix}
  {-1\over p+1} & -1 \\
  -1 & -(p+1) \\
 \end{pmatrix}
\ .
\label{bfive}
\end{equation}
That this second theory describes the transmutation of electrons of the same intrinsic charge but inverse vorticity (spin) can be easily recognised from the two Gauss laws for the statistical gauge field,
\begin{eqnarray}
2S_z^{+} + q^{+} &&= {1\over p+1} Q_{\alpha} \ , 
\nonumber \\
2S_z^{-} - q^{-} &&= {-1\over p+1} Q_{\alpha} \ ,
\label{addfive}
\end{eqnarray}
where $Q_{\alpha} = (1/2\pi) \int d^2 x \ \epsilon^{ij}\partial_i \alpha_j$ is the statistical emergent charge. The only simultaneous solution to these two equations when $q^{-}=  q^{+}$ requires $S_z^{-} = - S_z^{+}$.  As expected, the matrix $\Lambda^{-}$ has eigenvalues 0 and $\lambda^{-}=-\lambda^{+}$.

Exactly as one can group the two irreducible "flavours" of  two-components fermions into a four-component, reducible model without parity anomaly \cite{boy}, we can now also group the corresponding currents $(\phi^{\mu})^{+}$ and $(\phi^{\mu})^{-}$ into a single, enlarged model with charge current
\begin{equation}
j^{\mu } = (\phi^{\mu})^{+} - (\phi^{\mu})^{-} \ ,
\label{bbfive}
\end{equation}
and vortex current
\begin{equation}
\phi^{\mu } ={1\over 2} \left(  (\phi^{\mu})^{+} + (\phi^{\mu})^{-} \right) \ .
\label{bbbfive}
\end{equation}
This is given by
\begin{equation}
{\cal L} = {1\over 4\pi} {\cal A}^T_{\mu} \Lambda \ \epsilon^{\mu \nu \sigma} \partial_{\nu}{\cal A}_{\sigma} + 
{1\over 2\pi} {\cal A}^T_{\mu} \epsilon^{\mu \nu \sigma} \partial_{\nu}{\bf A}_{\sigma} \ ,
\label{bseven}
\end{equation}
with ${\cal A}_{\mu} = \left( \alpha _{\mu }, a^{+}_{\mu}, a^{-}_{\mu} \right)$ and ${\bf A} = \left( 0, A_{\mu}, -A_{\mu} \right)$ and
\begin{equation}
\Lambda = 
\begin{pmatrix}
  0 & -1 &  -1 \\
  -1 & p+1 & 0 \\
  -1 & 0 & -(p+1) \\
\end{pmatrix}
\ .
\label{beight}
\end{equation}
The elements of $\Lambda$ in the first column and row are invariant under P and T transformations, since they involve the mixed Cherns-Simons coupling of a vector gauge field with a pseudovector one. The remaining diagonal elements, when summed in the Lagrangian, represent also the mixed Chern-Simons term between a vector and a pseudo vector gauge fields. The whole model (\ref{beight}) is thus P- and T-invariant, as anticipated.  Correspondingly the eigenvalues of the matrix $\Lambda$ are given by 0, corresponding to the single superfluid mode and by two equal and opposite values $\pm \lambda$, with 
\begin{equation}
\lambda = \sqrt{(p+1)^2 +2} \ .
\label{bbeight}
\end{equation}
The possibility of obtaining a P- and T-invariant doubled anyon system by combining the two spin states of 2D fermions (after breaking the spin symmetry) was already pointed out in \cite{weiss}, where also the relevance of pseudovector statistics-changing gauge fields was suggested. 

The coordinate transformation that diagonalizes the model (\ref{bseven})  is given by
\begin{eqnarray}
\cal{A}_{\mu} &&= O \ \cal{W}_{\mu} \ , \qquad \quad \ \cal{W}_{\mu} = (\varphi_{\mu}, \omega^{+}_{\mu}, \omega^{-}_{\mu}) \ ,
\nonumber \\
\Lambda_D &&= O^T \Lambda \ O \ , \qquad \Lambda_D = \begin{pmatrix}
  0 & 0 &  0  \\
  0 & \lambda^3 & 0 \\
  0 & 0 & -\lambda^3  \\
\end{pmatrix} \ ,
\nonumber \\
O &&= \begin{pmatrix}
  p+1 & -1  &  -1  \\
  1 &  {p+1\over 2} + {1\over 2} \lambda &  {p+1\over 2} - {1\over 2} \lambda\\
  -1  & -{p+1\over 2} + {1\over 2} \lambda & -{p+1\over 2} - {1\over 2} \lambda  \\
\end{pmatrix} \ ,
\label{bnine}
\end{eqnarray}
where the columns of $O$ express the diagonal fields $(\varphi_{\mu}, \omega^{+}_{\mu}, \omega^{-}_{\mu})$ as linear combinations of the original fields $\left( \alpha _{\mu }, a^{+}_{\mu}, a^{-}_{\mu} \right)$ (note that $O$ is not an orthogonal matrix: its rows and columns are orthogonal but we have chosen a length $\lambda^2$ instead of 1 to better expose the physical content of the model). In the new variables, the doubled anyon superfluidity model has the Lagrangian density
\begin{eqnarray}
{\cal L} &&= {1\over 4\pi} {\cal W}^T_{\mu} \Lambda_D \ \epsilon^{\mu \nu \sigma} \partial_{\nu}{\cal W}_{\sigma} + 
{2\over 2\pi}  \varphi_{\mu }\epsilon^{\mu \nu \sigma} \partial_{\nu}A_{\sigma} + 
\nonumber \\
&&+ {p+1\over 2\pi} \omega^{+}_{\mu}\epsilon^{\mu \nu \sigma} \partial_{\nu}  A_{\sigma}
+ {p+1\over 2\pi}  \omega^{-}_{\mu} \epsilon^{\mu \nu \sigma} \partial_{\nu} A_{\sigma }\ . 
\label{bten}
\end{eqnarray}
Note that the gapless mode gauge field $\varphi_{\mu}$ is a pseudovector under  P and T so that the corresponding current describes a scalar particle carrying intrinsic charge 2 and emergent charge $(p+1)$. The massive modes, instead combine into a charge $(p+1)$, 2D vector particle with Chern-Simons mass: this has two degrees of freedom, each carrying both charge and vorticity, that are interchanged under P and T. There is no scalar Higgs boson. At this point one can as well forget about the two original spin components in the construction of the doubled model: (\ref{bten}) stands as a P- and T-invariant "anyon" superfluidity model per se. 

Let us now switch from superfluidity to superconductivity. In order to illustrate how the photon acquires mass we shall consider only the low-energy sector of the model, at energies well below the mass of the vector $\omega^{\pm}_{\mu}$, so that this is frozen to all purposes. Including the Maxwell action for the photon and the dynamical term for the gapless mode (first term in a derivative expansion) we obtain
\begin{equation}
{\cal L}  =  -{1\over 4 e^2 } F_{\mu \nu}F^{\mu \nu} + {2\over 2 \pi} \varphi_{\mu} \epsilon^{\mu\nu\sigma}\partial_\nu A_\sigma -{1\over 4 g^2 } f_{\mu \nu } f^{\mu \nu}  \ ,
\label{beleven}
\end{equation}
where $F_{\mu \nu} = \partial_{\mu}A_{\nu} - \partial_{\nu} A_{\mu}$, $f_{\mu \nu} = \partial_{\mu}\varphi_{\nu} - \partial_{\nu} \varphi_{\mu}$ (for simplicity of presentation we consider the relativistic version of the model) and $g^2$ is a coupling constant with dimension mass. Note that the Maxwell term for $\varphi_{\mu}$ is the unique gauge invariant (and relativistic) term that can appear at $O(1/m)$ in a derivative expansion. The equations of motions for this model are
\begin{eqnarray}
\partial_\mu F^{\mu\nu} &&= - {e^2 \over \pi} \epsilon^{\nu \alpha\beta} \partial_{\alpha} \varphi_{\beta} \ , \nonumber \\
\partial_\mu f^{\mu\nu}  &&= - {g^2 \over \pi} \epsilon^{\nu \alpha\beta} \partial_{\alpha} A_{\beta}  \ ,
\label{btwelve}
\end{eqnarray}
which can be easily combined \cite{jac} to give
\begin{equation}
\left[ \Box + m^2 \right] F_{\mu\nu} = 0 \ ,
\label{bthirteen}
\end{equation}
with $m=eg/\pi$, which shows that the photon has become massive and electromagnetic fields are screened. This screening is the anticipated Meissner effect. Let us call
\begin{equation}
J^{\mu} = {1\over 2\pi} \epsilon^{\mu \nu \sigma} \partial_{\nu} \varphi_{\sigma} = j^{\mu} +{p+1 \over 2\pi} \epsilon^{\mu \nu \sigma} \partial_{\nu} \alpha_{\sigma} \ ,
\label{bfourteen}
\end{equation}
the current of the gapless mode carrying intrinsic charge 2 and emergent charge $(p+1)$. Expressing this as $J^{\mu} = (1/4\pi) \epsilon^{\mu \alpha \beta} f_{\alpha \beta}$ and inserting this into the second equation (\ref{btwelve}) we obtain
\begin{equation} 
\epsilon^{\mu \alpha \beta} \partial_{\alpha} J_{\beta} = {g^2\over 2\pi} \epsilon^{\nu \alpha \beta} F_{\alpha \beta} \ ,
\label{bfifteen}
\end{equation}
or, in components (for $\partial_i J^0=0$)
\begin{eqnarray}
\partial_t {\bf J} &&= {g^2\over \pi} {\bf E} \ ,
\nonumber \\
\nabla \times {\bf J} &&= {g^2\over \pi} B \ .
\label{bsixteen}
\end{eqnarray}
These are the London equations, which state that the combination of 2 intrinsic charges with (p+1) emergent charges can flow without resistance. 

\section{Mutual anyon superconductivity}
In this section we shall show that the Higgsless superconductivity mechanism described above takes a particularly simple and familiar form in another set of coordinates, one that can be easily generalised to any number of dimensions. 
 
 To this end let us start from the original formulation (\ref{bseven}) and (\ref{beight}) and apply the coordinate transformation 
 \begin{eqnarray}
\cal{A}_{\mu} &&= O \ {\cal{M}}_{\mu} \ , \qquad \quad \ {\cal{M}}_{\mu} = (a_{\mu}, b_{\mu },  -\alpha_{\mu}) \ ,
\nonumber \\
\Lambda_M &&= O^T \Lambda \ O \ , \qquad \quad \quad O = \begin{pmatrix}
 0 & 0  &  -1  \\
 {1\over 2} & 1 &  0 \\
  {1\over 2}  & -1 & 0\\
\end{pmatrix} \ .
\label{cone}
\end{eqnarray}
In these coordinates the Lagrangian density of the model becomes (after an integration by parts in the action for the electromagnetic coupling) 
\begin{equation} 
{\cal L} = {1\over 4\pi} {\cal M}^T_{\mu} \Lambda_M \ \epsilon^{\mu \nu \sigma} \partial_{\nu}{\cal M}_{\sigma} + 2 A_{\mu } j^{\mu} \ ,
\label{ctwo}
\end{equation}
with 
\begin{equation}
\Lambda_M= \begin{pmatrix}
  0 & p+1 &  1  \\
  p+1 & 0 & 0 \\
  1 & 0 & 0  \\
\end{pmatrix} \ .
\label{cthree}
\end{equation}
It involves only mutual Chern-Simons terms,
\begin{equation}
{\cal{L}} = {p+1\over 2\pi} a_{\mu} \epsilon^{\mu \nu \sigma} \partial_{\nu} b_{\sigma} + {1\over 2\pi} a_{\mu} \epsilon^{\mu \nu \sigma} \partial_{\nu} \alpha_{\sigma} + 2 A_{\mu} j^{\mu} \ .
\label{ctwo}
\end{equation} 
and describes specifically a mutual Chern-Simons interaction between vortices with conserved current 
\begin{equation}
\phi^{\mu} ={1\over 2} \left(  (\phi^{\mu})^{+} + (\phi^{\mu})^{-} \right) = {1\over 2\pi} \epsilon^{\mu \nu \sigma} \partial_{\nu} a_{\sigma} \ ,
\label{cthree}
\end{equation}
and charges with conserved current 
\begin{equation}
j^{\mu} = (\phi^{\mu})^{+} - (\phi^{\mu})^{-})= {1\over 2\pi} \epsilon^{\mu \nu \sigma} \partial_{\nu} b_{\sigma} \ ,
\label{cfour}
\end{equation}
and charge unit 2. In addition there is also a mutual Chern-Simons interaction between the vortices and the statistical gauge field. Since $a_{\mu}$ and $b_{\mu}$ are vector and pseudovector gauge fields, respectively and the statistical gauge field $\alpha_{\mu}$ is also a pseudovector, this Lagrangian respects both the discrete symmetries P and T. 

In this formulation, the superconductivity mechanism is very simple to illustrate. In the average field approximation, the statistical gauge fields provides a uniform emergent charge for the vortices. Keep in mind that this emergent charge is different from the intrinsic charge that couples to $A_{\mu}$. Due to the Magnus force, which is the exact dual of the Lorentz force in 2D, 
vortices fall into Landau levels and any configuration with a completely filled Landau level has a gap. This is valid, in particular, for the vacuum with all Landau levels empty, i.e. the state with no vortices. This gap for vortices represents the Meissner effect. Generic intrinsic charge fluctuations are also gapped since, due to the mutual Chern-Simons term, they distort the uniform emergent background charge field for the vortices, and must thus be accompanied by vortex or anti-vortex nucleation, exactly the same mechanism as in traditional anyon superconductivity \cite{chen}. There is however a particular combination of 1 intrinsic charge (with charge unit 2) and $(p+1)$ emergent charges that can fluctuate freely without altering the overall charge distribution felt by the vortices: this is the isolated gapless mode leading to superfluidity. This implies that low-energy emergent charges carry intrinsic charge $2/(p+1)$. The relative Aharonov-Bohm statistical phase acquired by one such emergent charge when it is carried around a vortex is thus $\pi (p+1)$. This is compatible with the relative statistical phase $\pi $ of one intrinsic charge 2 only if $p$ is an even integer. In other words, only if $p$ is an even integer, the emergent charge can be attached to intrinsic charges with charge unit 2 without altering the real flux quantization in units of $\pi$. 

In these variables the relation to the original mechanism of anyon superconductivity becomes fully exposed. If we freeze fluctuations of the emergent gauge field $\alpha$ in (\ref{ctwo}) we are left with the average field approximation in which both vortices and intrinsic charges are gapped and with the effective action given by the first mutual Chern-Simons term in (\ref{ctwo}). For $p=0$ this is nothing else than the effective action for a 2D topological insulator \cite{moore}, which describes exactly a state in which both vortices and charges are fully screened. While the gapped average field state of anyon superconductivity was an incompressible quantum Hall state, in the present model it is a topological insulator. The additional gauge field that leads to an isolated superfluid mode in anyon superconductivity was the statistics changing Chern-Simons field, carrying vorticity and thus breaking the P and T symmetries. In the present mode, the mechanism leading to an isolated superfluid mode is exactly the same: the additional gauge field however is a pseudovector that has its origin in the condensation of topological defects, as we show below. It carries a scalar emergent charge and thus does not break P ans T. 

Having discussed in detail how the gapless superfluid mode arises, let us now turn to the effects of the electromagnetic coupling in (\ref{ctwo}). We have already seen in the previous section that this causes the entire model to become gapful via the BF topological mass mechanism \cite{bow}. This, on the other side, goes hand in hand with topological order \cite{wen}, characterised by a ground state degeneracy $d^g$ on  Riemann surfaces of genus $g$. The degeneracy parameter $d$ for multi-component Chern-Simons terms $(1/4\pi) {\bf a}_{\mu} M \epsilon^{\mu \nu \sigma} \partial_{\nu} {\bf a}_{\sigma}$  is governed by the determinant of $M$ \cite{poly}. If $M$ has only integer entries, then $d=|{\det } \ M|$. If the entries are rational one has to construct a representation $M=M_1 M_2^{-1}$ where $M_1$ and $M_2$ have mutually prime integer entries. The degeneracy parameter is then $d=|{\det } \ M_1 \ {\rm det}\ M_2|$. 

Using (\ref{cfour}) one can formulate (\ref{ctwo}) as a unique multicomponent Chern-Simons model
\begin{eqnarray}
{\cal L} &&= {1\over 4\pi} {\cal J}^T_{\mu} Q \ \epsilon^{\mu \nu \sigma} \partial_{\nu}{\cal J}_{\sigma} \ ,
\nonumber \\
Q &&= \begin{pmatrix}
  0 & 0 &  2 & 0 \\
  0 & 0 &  p+1 & 1  \\
  2 & p+1 & 0 & 0   \\
  0 & 1 & 0 & 0\\
\end{pmatrix} \ ,
\label{csix}
\end{eqnarray}
for ${\cal{J}}_{\mu}=\left( A_{\mu}, a_{\mu}, b_{\mu}, -\alpha_{\mu}\right)$. Since all entries of $Q$ are integer and ${\rm det\ Q} = 4$, this implies topological order with degeneracy parameter $d=4$. 

To complete the model we should also give a picture of how superconductivity is destroyed at temperatures approaching the high-energy vector mass. To this end we recall that the mass of the vector particle is of the order of the gap for vortices in the average field approximation. At temperatures comparable with this energy scale vortices can cross this gap and become thus liberated. We expect thus the transition to be due to the deconfinement of vortices, which, in 2D is a Kosterlitz-Thouless transition \cite{kos} and, as expected is not related to the restoration of a symmetry, nor does it involve an order parameter. This is fully confirmed experimentally. Indeed, in 2D, our model of topological Higgsless superconductivity is explicitly realized \cite{dst2} as global superconductivity in Josephson junction arrays \cite{faz}. In these granular materials, superconductivity is known to be destroyed at high temperatures due to the deconfinement of vortices in a Kosterlitz-Thouless phase transition \cite{faz}: no spontaneous symmetry breaking is involved. 

Let us recapitulate the properties of this novel superconductivity mechanism. At the superfluid level there is one isolated gapless mode of charge 2 but no Higgs boson; rather, there is a charged, massive Chern-Simons vector. When electromagnetism is switched on, the photon acquires mass by combining with the gapless mode via the topological BF mass mechanism \cite{jac, bow}. The resulting topological order is 4. Although the charge 2 of the gapless mode and the topological order $d=4$ coincide with those of the abelian Higgs model \cite{han}, this is quite different than the standard BCS pairing mechanism. Here, the individual components of a charge 2 "pair" are already bosons, having been statistically transmuted from fermions by an emergent statistical gauge field. Nor can this model be phenomenologically described by the Abelian Higgs model: there is no local order parameter and, correspondingly, there is no Higgs boson. The properties of the present model are, rather suggestive of a possible connection with Higgsless models of spontaneous symmetry breaking via extra dimensions and the AdS/CFT correspondence \cite{csa}. 

Also note that this model is a paradigm example of the fact that one cannot deduce the characteristics of a superconductivity model by looking at the low-energy (photon plus gapless mode) sector exclusively. 
Doing so in this case would have lead to the erroneous conclusion that this model coincides with the abelian Higgs model, since both the charge of the gapless mode and the topological order are the same. Instead, in order to establish the nature of a superconductivity model it is crucial to focus on the mechanism by which the gap is opened and the gapless mode is originated and this is intimately connected with the high-energy sector. In a sense, all possible superconductivity models must have the same phenomenological low-energy structure, it is only the origin of the gap, and thus the high-energy physics, that can distinguish them. In the present case, the gapless mode is intimately tied to the emergent statistical gauge field. We must thus clarify the origin of this gauge field. This is the subject of the next section. 

\section{The emergent gauge field as a topological excitation}
The mutual Chern-Simons interaction between charges and vortices, first term in equation (\ref{ctwo}), is easily generalisable \cite{dst2} to any number of dimensions as a BF term \cite{birmi},
\begin{equation}
S_{BF} = {k \over 2 \pi} \int_{M_{d+1}} a_1 \wedge d b_{d-1} \ ,
\label{done}
\end{equation}
on a manifold of spatial dimension $d$. Here $a_1$ is a one-form and, correspondingly, $b_{d-1}$ is a $(d-1)$-form. The conserved current $j_1 = * db_{d-1}$ represents charge fluctuations, while the generalized current $\phi_{d-1} = * da_1$ describes conserved fluctuations of $(d-2)$-dimensional branes. The form $b_{d-1}$ is a pseudo-tensor, while $a_1$ is a vector: the BF coupling is thus P- and T-invariant. The BF term always represents a mass term for the gauge fields $a_1$ and $b_{d-1}$ \cite{dst2}. In the special case $d=2$, it reduces exactly to the mutual Chern-Simons term in (\ref{ctwo}). 

The action (\ref{done}) has the usual gauge symmetry under shifts 
\begin{equation}
a_1\rightarrow a_1+ \xi_1 \ ,
\label{dtwo}
\end{equation}
with $\xi_1$ a closed 1-form, $d \xi_1 = 0$, provided vanishing boundary conditions for the corresponding field strength are chosen. However, it has also a generalized Abelian gauge symmetry under transformations
\begin{equation}
b_{d-1} \rightarrow b_{d-1} + \eta_{d-1} \ ,
\label{dthree}
\end{equation}
where $\eta_{d-1}$ is a closed $(d-1)$-form: $d \eta_{d-1} = 0$. The important point is that, in application to low-energy effective models of condensed matter systems, these gauge symmetries have to be considered as {\it compact}. As is well known \cite{polya}, the compactness of the gauge fields leads to the presence of topological defects. In the present case there are both magnetic topological defects, associated with the compactness of the usual gauge symmetry (\ref{dtwo}) and electric ones, associated with the compactness of the gauge symmetry (\ref{dthree}). The electric topological defects  couple to the form $a_1$ and are string-like objects  described by a singular, closed 1-form $Q_1$: they describe the world lines of point charges. Magnetic topological 
defects couple to the form $b_{d-1}$ and are closed $(d-1)$-branes described  by a singular $(d-1)$-dimensional form $\Omega_{d-1}$. In 2 space dimensions they also reduce to string-like objects that describe the world lines of point vortices. 
These forms represent the singular parts of the field strenghts $da_1$ and $db_{d-1}$, allowed by the compactness of the gauge
symmetries \cite{polya}, and are such that the integral of their Hodge dual  over any  hypersurface of dimensions d and 2, respectively, is $2 \pi n$ with $n$ an integer.  In an effective field theory approach they have structure on the
scale of the ultraviolet cutoff. 

Thus far we have described the kinematics of topological defects. The dynamics depends, of course on all terms present in the full action. Concretely, however, topological defects can be dilute, in which case they do not have any effect, or can condense. The phase with condensed topological defects has a completely different character. In this section we shall not discuss the conditions for condensation of topological defects but we will instead focus on the characteristics of the phase in which the electric topological defects condense. 

A formal derivation of the action with condensed topological defects requires the introduction of an ultraviolet regularization, e. g. in the form of a lattice gauge theory. The result of this procedure \cite{polya}, however, amounts to promote the form $Q_1$ to a dynamical field over which one has to sum in the partition function, 
\begin{eqnarray}
&Z = \int {\cal D}a_1 {\cal D}b_{d-1} {\cal D}Q_1 \nonumber \\ 
&\exp \left[ i {k \over 2 \pi}\int_{M_{d+1}} \left( a_1 \wedge d b_{d-1} + a_1 \wedge *Q_1 \right) \right] \ .
\label{dfour}
\end{eqnarray}
Since $Q_1$ is closed, one can represent it as $Q_1 = d \alpha_{d-1}$. The summation over $Q_1$ in the partition function can then be substituted with a summation over $\alpha_{d-1}$ provided the gauge volume due to the additional symmetry $\alpha_{d-1} \to \alpha_{d-1} + \lambda_{d-1}$ with $\lambda_{d-1}$ a closed $(d-1)$ form, is duly subtracted. The resulting model in the electric condensation phase can thus be formulated in terms of the three dynamical gauge fields $a_1$, $b_{d-1}$ and $\alpha_{d-1}$ and has the action
\begin{equation}
S = {k \over 2 \pi}\int_{M_{d+1}} \left( a_1 \wedge d b_{d-1} + a_1 \wedge d\alpha_{d-1} \right)  \ .
\label{dfive}
\end{equation}
This is the generalization to any dimensions of the model (\ref{ctwo}) for the case $p=0$. In 2 space dimensions, $d=2$, it reduces exactly to the $p=0$ version of (\ref{ctwo}). The scalar gapless mode is represented by the gauge field combination $\varphi_{d-1}= b_{d-1}-\alpha_{d-1}$, with corresponding conserved current $j_1 = *d\varphi_{d-1}$. The remaining $d$ massive degrees of freedom in $a_1$ and $b_{d-1}+\alpha_{d-1}$ represent the massive vector in $d$ space dimensions. 

This shows that the origin of the emergent gauge field in this superconductivity model lies in the condensation of topological defects in effective, compact BF field theories of topological matter. It is no wonder that there is no Higgs boson in this model: there is no local order parameter, superconductivity arises as a consequence of a condensation of topological defects and the two phases can be distinguished only by the behaviour of Wilson loops \cite{polya,sim}. Notice that, due to the BF mass term, the topological defects $Q_1$ have short-range interactions: in a Euclidean formulation, their self-energy is proportional to the length of their world lines, exactly as their entropy. The condensation (or lack thereof) is thus determined by an energy-entropy balance. Since it arises due to the condensation of topological defects without any local order parameter, we find it appropriate to call this superconductivity model "topological". 

In two space dimensions this topological Higgsless superconductivity is explicitly realized \cite{dst1} as global superconductivity in Josephson junction arrays \cite{faz}. The 3D case is perhaps even more interesting, as we now show. 

\section{3D: turning a topological insulator into a superconductor} 
In three space dimensions the BF term (\ref{done}) is again the low-energy effective action for topological insulators \cite{moore}. These are topological states of matter in which both charges and vortices are completely screened in the bulk, but which support metallic edge states \cite{topins}. Our results imply that, if topological defects condense, topological insulators could turn into Higgsless topological superconductors described by the Lagrangian
\begin{equation}
{\cal{L}} = {1\over 2\pi} a_{\mu} \epsilon^{\mu \nu \sigma \rho} \partial_{\nu} b_{\sigma \rho} + {1\over 2\pi} a_{\mu} \epsilon^{\mu \nu \sigma \rho} \partial_{\nu} \alpha_{\sigma \rho} \ ,
\label{eone}
\end{equation} 
where we have left out, for simplicity, the electromagnetic coupling. Here $b_{\mu \nu}$ and the emergent gauge field $\alpha_{\mu \nu}$ are Kalb-Ramond antisymmetric (two-form) gauge fields \cite{kalb}, with generalized gauge invariance under the transformations
\begin{eqnarray}
b_{\mu \nu} &&\to b_{\mu \nu} + \partial_{\mu} \lambda_{\nu}-\partial_{\nu} \lambda_{\mu} \ ,
\nonumber \\
\alpha_{\mu \nu} &&\to \alpha_{\mu \nu} + \partial_{\mu} \eta_{\nu}-\partial_{\nu} \eta_{\mu} \ .
\label{etwo}
\end{eqnarray}
The dual field strength 
\begin{equation}
j^{\mu}= {1\over 2\pi} \epsilon^{\mu \nu \sigma \rho} \partial_{\nu}b_{\sigma \rho} \ ,
\label{ethree}
\end{equation}
of the Kalb-Ramond field is a vector field (since the emergent gauge field is a pseudotensor) which represents the charge current, while the current of vortex strings is given by the usual antisymmetric dual field strength
\begin{equation}
\phi^{\mu \nu} = {1\over 2\pi } \epsilon^{\mu \nu \sigma \rho} \partial_{\sigma} a_{\rho} \ . 
\label{efour}
\end{equation}

The superconductivity mechanism in three dimensions parallels exactly the 2D mutual anyon superconductivity presented above, as we now show.  In the average field approximation, the emergent gauge field provides a uniform emergent charge given by the Kalb-Ramond dual field strength $F^0 = \epsilon^{ijk} \partial_i \alpha_{jk}$ . This uniform Kalb-Ramond emergent charge causes a gap for vortices via the Magnus force. To see this, let us consider an elementary vortex string with world-surface parametrized by ${\bf x}(\xi_0, \xi_1)$ and action
\begin{equation}
S_v=\int d^2 \xi  \ T\ \sqrt{g} g^{ab} {\cal D}_a x_{\mu} {\cal D}_b x^{\mu} + \alpha_{\mu \nu}  \epsilon^{ab} \partial_a x^{\mu} \partial_b x^{\nu}  \ ,
\label{esix}
\end{equation}
where ${\cal{D}}_a$ are the covariant derivatives with respect to the induced metric $g_{ab}=\partial_a x_{\mu} \partial_b x^{\mu}$, $g$ is the determinant of this induced metric and $2T$ is the (bare) string tension. The first term in this action is the celebrated Polyakov action \cite{polya} whereas the second term represents the Magnus force coupling to the antisymmetric Kalb-Ramond emergent gauge field. 

We analyze this model along the lines of \cite{klei} by introducing a Lagrange multiplier $l^{ab}$ to enforce the constraint $g_{ab}= \partial_a x_{\mu} \partial_b x^{\mu}$ and extending the action (\ref{esix}) to 
\begin{equation}
S_v \to S_v + \int d^2 \xi \sqrt{g} \ l^{ab} \left( \partial_a x_{\mu} \partial_b x^{\mu} - g_{ab} \right) \ . 
\label{eseven}
\end{equation}
We then parametrize the world-surface in a Gauss map by choosing to set the coordinate "1" of space-time along the vortex, $x_{\mu} (\xi_0, \xi_1) = \left( \xi_0, \xi_1, \phi^i (\xi_0, \xi_1) \right)$, where $\phi^i (\xi_0, \xi_1)$, $i=2,3$, describe the 2 transverse fluctuations. With the usual homogeneity and isotropy ansatz $g_{ab} = \rho \ \eta_{ab}$, $l^{ab} = l \ g^{ab}$ 
we obtain
\begin{equation}
S=\int d^2 \xi \ \theta +  \int d^2 \xi  \ {T_r\over 2} \partial_a \phi^i \partial^a\phi^i + \alpha_{\mu \nu}  \epsilon^{ab} \partial_a x^{\mu} \partial_b x^{\nu}
\label{eeight}
\end{equation}
where $\theta =T_r -2l\rho$ and $T_r = 2(T+l)$ is the renormalized string tension. At this point we can partially fix the gauge for the antisymmetric Kalb Ramond gauge field by choosing the partial Weyl gauge conditions $\alpha_{02} = 0$, $\alpha_{03} = 0$ and the partial axial gauge condition $\alpha_{23}=0$. This gives 
\begin{equation}
S=\int d^2 \xi \ {T_r\over 2} \ \partial_a \phi^i \partial^a\phi^i - \theta_0 - \theta_i \dot \phi^i \ ,
\label{enine}
\end{equation}
where we have omitted the first constant term (irrelevant for the following) and 
\begin{equation}
\theta _0 = \alpha_{10}\ , \qquad \theta_i = \alpha_{1i} \ , \qquad i=2,3 \ , 
\label{eten}
\end{equation}
are the components of an effective $(2+1)$-dimensional gauge field with residual gauge invariance under transformations 
\begin{eqnarray}
\theta_0 &&\to \theta_0 - \partial_0 \lambda_1 \ ,
\nonumber \\
\theta_i && \to \theta_i - \partial_i \lambda_1 \ ,
\label{eeleven}
\end{eqnarray}
where $\lambda_1$ is the first component of the original vector gauge parameter of the two-form Kalb-Ramond gauge field, which embodies the only residual gauge freedom in this gauge.  The Hamiltonian corresponding to (\ref{enine}) is 
\begin{equation}
H = \int d\xi_1 \ {1\over 2T_r} \left(  P^i-  \theta^i \right)^2 + {T_r\over 2} \partial_{\xi_1}\phi^i \partial_{\xi_1}\phi^i  + \theta_0 \ ,
\label{etwelve}
\end{equation}
where $P^i = T_r \dot \phi^i + \theta^i$ is the canonical momentum density. This Hamiltonian is equivalent to a continuous sequence labeled by $\xi_1$ of particles, held together by the elastic term ${T_r\over 2} \partial_{\xi_1}\phi^i \partial_{\xi_1}\phi^i$ and subject to an effective $(2+1)$-dimensional electromagnetic potential $\theta_{\mu}$. In the gauge we have chosen, a uniform Kalb-Ramond dual field strength $F^0 = \epsilon^{ijk} \partial_i \alpha_{jk}$ is equivalent to a uniform, effective $(2+1)$-dimensional emergent charge (dual magnetic field) $F^0 = \partial_3 \theta_2 - \partial_2 \theta_3$. Each "particle" in the sequence constituting the vortex feels thus a uniform emergent charge due to the Magnus force. The ground state energy of a vortex is thus given by the sum of the zero-point energies of a sequence of Landau oscillators, $E = (F^0/2T_r) (L/\zeta)$, where $L$ is the length of the vortex and $\zeta$ the ultraviolet cutoff. As a consequence, vortices are gapped in presence of a uniform Kalb-Ramond dual field strength (emergent charge): this gap leads to the Meissner effect. 

Exactly as in the 2D case, generic intrinsic charge fluctuations are also gapped, since a generic fluctuation in the current (\ref{ethree}) causes a distortion in the total charge distribution felt by the vortices, with the consequent nucleation of vortices or anti-vortices, which costs energy, as we have just shown above. There is however one particular combination $\varphi_{\mu \nu} = b_{\mu \nu} - \alpha_{\mu \nu}$ of intrinsic and emergent charges that can oscillate coherently without modifying the total background charge distribution felt by the vortices: this is the isolated gapless mode implying superfluidity in the model (\ref{eone}) and superconductivity when it is coupled to electromagnetic fields. The remaining 3 degrees of freedom $a_{\mu}$ and $b_{\mu \nu} + \alpha_{\mu \nu}$ are coupled by a topological BF term as is evident from (\ref{eone}). As is well known \cite{bow} they represent thus a massive vector (spin 1) particle. Also in this case there is no scalar Higgs boson and no local order parameter. 

As we have already pointed out, the first term in (\ref{eone}) is the low-energy effective field theory for topological insulators \cite{moore}. As in 2D, thus, the average field state for our superconductivity mechanism is a fully gapped topological insulator. Actually, our results imply that a 3D topological insulator is the 3D analogue of a full first Landau level when a Kalb-Ramond scalar density $F^0$ is applied to vortex strings instead than a 2D pseudoscalar magnetic field to charged particles. 
Our results imply, thus, that the condensation of the emergent gauge field $\alpha_{\mu \nu}$ could thus turn a topological insulator into a topological Higgsless superconductor. But what could drive this phase transition?  The original idea of formulating low-energy effective field theories of condensed matter systems in terms of emergent gauge fields was based on the fact that, typically, the dynamical terms for the gauge fields are infrared irrelevant and, thus, the whole physics is determined by the topological terms. First of all, in 3D, the usual Maxwell term for the gauge field $a_{\mu}$ is marginal and must anyhow be included in the action. Secondly, as is well known \cite{kog}, when the theory is considered as a cutoff theory, as it must in the present context,  this argument is valid only in the perturbative scaling regime in which all masses of irrelevant operators are beyond the ultraviolet cutoff. In the opposite regime, however, when there are masses of high-energy fields in the observable regime below the cutoff, these can induce non-perturbative effects, like phase transitions. This is exactly what happens in this model, as we now show. 

Let us do so in the (relativistic) Euclidean space formulation of the model with the dynamical gauge field terms added to the topological BF term
\begin{eqnarray}
S_E &&= \int d^3 x {1\over 4e^2} f_{\mu \nu} f_{\mu \nu} -  {i\over 2\pi} a_{\mu} \epsilon^{\mu \nu \sigma \rho} \partial_{\nu} b_{\sigma \rho} +
\nonumber \\
&&+ {1\over 12g^2} h_{\mu \nu \alpha } h_{\mu \nu \alpha} - {i\over 2\pi} a_{\mu} Q_{\mu}\ , 
\label{ethirteen}
\end{eqnarray}
where we have included the topological defects $Q_{\mu}$ due to the compactness of the $BF$ terms, which in Euclidean space become strings corresponding to the Minkowski world lines of charges and 
\begin{eqnarray}
f_{\mu \nu} &&= \partial_{\mu}a_{\nu}-\partial_{\nu}a_{\mu} \ ,
\nonumber \\
h_{\mu \nu \sigma} &&= \partial_{\mu } b_{\nu \sigma} + \partial_{\nu } b_{\sigma \mu} + \partial_{\sigma } b_{\mu \nu} \ .
\label{efourteen}
\end{eqnarray}
The first term in (\ref{ethirteen}) is the usual Maxwell term for the gauge field $a_{\mu}$, the third one is the dynamical term for the three-form Kalb-Ramond field strength $h_{\mu \nu \sigma}$. Recalling that the dual Kalb-Ramond field strength $f_{\mu} = (1/6) \epsilon_{\mu \nu \sigma \rho} \partial_{\nu}b_{\sigma \rho} = \pi j_{\mu}/3$ coincides with the conserved charge current, eq. (\ref{ethree}), we see that the Kalb-Ramond dynamical term represents a charge-charge interaction with strength $g$ of dimension mass. Correspondingly, since $\tilde f_{\mu \nu} = 4\pi \phi_{\mu \nu}$ represents the conserved vortex current, eq. (\ref{efour}), the Maxwell term embodies a vortex-vortex interaction with dimensionless strength $e$. 

We can now integrate out the gauge fields $a_{\mu}$ and $b_{\mu \nu}$ to obtain en effective action for the topological excitations $Q_{\mu}$. As we have explained in the previous section, however, a proper treatment of the topological excitations requires the introduction of an ultraviolet cutoff, by formulating the model, e.g. on a lattice. A fully self-contained derivation of the lattice effective action for the topological excitations is beyond the scope of the present paper. We have presented the complete calculation elsewhere \cite{dst1, dst3}, here we simply quote the result, which is the evident lattice translation of the result one would obtain from the continuous action (\ref{ethirteen}). The partition function is given by
\begin{eqnarray}
Z_{\rm top} &&= \sum_{\left\{ Q_{\mu} \right\}} {\rm exp} \left( -S_{\rm top} \right) \ ,
\nonumber \\
S_{\rm top} &&= \sum_{{\bf x}} {e^2\over 2 l^2} \ Q_{\mu} {\delta_{\mu \nu} \over m^2-\nabla^2} Q_{\nu} \ ,
\label{efifteen}
\end{eqnarray}
where $l$ is the lattice spacing  and $m=eg/\pi$ is the topological BF mass. This mass causes the screening of both charges and vortices in the topological insulator phase. As expected, the topological excitations have also short-range interactions on the scale of this mass. The phase structure implied by this result can be derived by the usual free energy arguments \cite{kle2}.
By approximating the short-range interactions with contact terms one can assign an energy $ (e^2/2l^2m^2) L$ to a topological excitation of length $L$. The entropy of a string of length $L$ is also proportional to $L$, $\mu L$, where $\mu \simeq {\rm ln} 7$ since, in 4 Euclidean dimensions a string has 7 directions to choose from, without backtracking. The free energy of a string of length $L$ is thus given approximately by
\begin{equation}
F = \left( {e^2\over 2 (lm)^2} - \mu \right) \ L \ .
\label{esixteen}
\end{equation}
When $ml < e/\sqrt{2\mu}$ the free energy is positive, hence it is minimised by strings of length 0 or, in other words, topological excitations are dilute: this is the topological insulator phase. If instead $ml > e/\sqrt{2\mu}$ the free energy is dominated by the entropy and becomes negative, hence long strings are favoured and topological excitations condense: this is the superconductor phase. This shows that, in a fully non-perturbative treatment including topological excitations, topological insulators develop a transition to Higgsless topological superconductivity when the range of the screened Coulomb interaction becomes smaller than a critical value $(\sqrt{2\mu}/e) l$. 

\section{Conclusions}
The anyon superconductivity mechanism, in its P- and T-invariant doubled formulation, can be extended to any number of dimensions. In 3D it involves a compact topological BF term between a usual vector gauge field and a two-form Kalb-Ramond gauge field. The emergent gauge field is also a two-form gauge field arising from the condensation of topological defects. The basic mechanism is the same as in 2D: in the average field approximation a uniform Kalb-Ramond emergent charge causes a gap for vortex strings via the Magnus force. The gapped average field state is a topological insulator. One particular combination of intrinsic and emergent charges, however, is gapless and leads to superfluidity. There is no local order parameter and, thus no Higgs scalar, rather the massive mode is a charged vector (spin 1) particle. This mechanism predicts a possible phase transition from topological insulators to Higgsless superconductors if the charge screening in the topological insulator becomes strong enough.


\begin{references}
\bibitem{wil}For a review see: F. Wilczek, {\it Fractional Statistics and Anyon Superconductivity}, World Scientific, Singapore (1990).
\bibitem{jac}R. Jackiw and S. Templeton, {\it Phys. Rev.} {\bf D23} (1981) 2291; 
S. Deser, R. Jackiw and S. Templeton, {\it Phys. Rev. Lett.} {\bf 48} (1982) 975;  {\it Ann. Phys.} (N.Y.) {\bf 140} (1982) 372.
\bibitem{gid}S. B. Giddins and F. Wilczek, {\it Mod. Phys. Lett. } {\bf A5} (1990) 635. 
\bibitem{topins}For a review see: M.Z. Hasan and C. L. Kane, {\it Rev. Mod. Phys} {\bf 82} (2010) 3045; M. Z. Hasan and J. E. Moore, {\it Ann. Rev. Cond. Matt. Phys.} {\bf 2} (2010) 55 AOP. 
\bibitem{zee}X.-G. Wen and A. Zee, {\it Phys. Rev.} {\bf B44} (1991) 274.
\bibitem{moe}R. Moessner and S. L. Sondhi, {\it Phys. Rev. Lett.} {\bf 105} (2010) 166401. 
\bibitem{magnus}R. Fazio, A. van Otterlo and G. Sch\"on, {\it Helv. Phys. Acta} {\bf 65} (1992) 228; Y. Kim and K. Lee, {\it Phys. Rev.} {\bf D49} (1994) 2941; X.-M. Zhu and B. Sundqvist, arXiv:cond-mat/9703072; M. Stone, arXiv:cond-mat/9708017. 
\bibitem{chen}Y.-H. Chen, F. Wilczek, E. Witten and B. Halperin, {\it Int. J. Mod. Phys.} {\bf 3} (1989) 1001. 
\bibitem{wen}X.-G. Wen and A. Zee, {\it Phys. Rev. Lett.} {\bf 62} (1989) 1937; {\it Phys. Rev.} {\bf B41} (1990) 240; {\it Int. J. Phys.} {\bf B4} (1990) 437; X.-G Wen, {\it Int. J. Mod. Phys.} {\bf B5} (1991) 1641. 
\bibitem{birmi}For a review see: D. Birmingham, M. Blau, M. Rakowski and G. Thompson, {\it Phys. Rep.} {\bf 209} (1991) 129. 
\bibitem{kalb}M. Kalb and P. Ramond, {\it Phys. Rev.} {\bf D9} (1974) 2273; V. I. Ogievetski and I. V. Polubarinov, {\it Sov. J. Nucl. Phys.} {\bf 4} (1967) 156. 
\bibitem{bow}T. J. Allen, M. Bowick and A. Lahiri, {\it Mod. Phys. Lett.} {\bf A6} (1991) 559; A. P. Balachandran and P. Teotonio-Sobrinho, {\it Int. J. Mod. Phys.} {\bf A8} (1993) 723; M. Bergeron, G. Semenoff and R. J. Szabo, {\it Nucl. Phys.} {\bf B437} (1995) 695. 
\bibitem{moore} G. Y. Cho and J. E. Moore, {\it Ann. Phys.} {\bf 326} (2011) 1515.
\bibitem{dst2}M. C. Diamantini, P. Sodano and C. A. Trugenberger {\it Eur. Phys. J.} {\bf 53} (2006) 19. 
\bibitem{dst1}M. C. Diamantini, P. Sodano and C. A. Trugenberger,  {\it  Nucl. Phys.} {\bf B474} (1996) 641.
\bibitem{faz}For a review see: Y. M. Blanter, R. Fazio and G. Schon, {\it Nucl. Phys. Proc. Supp.} {\bf 58} (1997) 79. 
\bibitem{kog}For a review see, e.g.: J. B. Kogut, {Rev. Mod. Phys.} {\bf 51} (1979) 659. 
\bibitem{dst3}M. C. Diamantini, P. Sodano and C. A. Trugenberger, {\it New. J. Phys.} {\it 14} (2012) 063013.
\bibitem{topsup}For a review see, e.g.: X.-L. Qi and S.-C Zhang, {\it Rev. Mod. Phys.} {\bf 83} (2011) 1057. 
\bibitem{han}T. H. Hansson, V, Oganesyan and S. L. Sondhi, {\it Ann. Phys. } {\bf 313} (2004) 497. 
\bibitem{csa}For a review see: C. Csaki, J. Hubisz and P. Meade, arXiv:hep-ph/0510275. 
\bibitem{jain}J. K. Jain, {\it Phys. Rev. Lett.} {\bf 63} (1989) 199, {\it Phys. Rev.} {\bf B40} (1989) 8079. 
\bibitem{boy}D. Boyanovski, R. Blankenbecler and R. Yahalom, {\it Nucl. Phys.} {\bf B270} (1986) 483.
\bibitem{hh}F. D. M. Haldane, {\it Phys. Rev. Lett.} {\bf 51} (1983) 605; B. I. Halperin, {\it Phys. Rev. Lett.} {\bf 52} (1984) 1583. 
\bibitem{weiss}G. F. Semenoff and N. Weiss, {\it Phys. Lett.} {\bf B250} (1990) 117. 
\bibitem{poly}A. Polychronakos, {\it Phys. Lett.} {\bf B241} (1989) 37.
\bibitem{kos}For a review see, e.g.: C. Itzykson and J.-M. Drouffe, {\it Statistical Field Theory}, Cambridge University Press, Cambridge (1989). 
\bibitem{polya}A. Polyakov, {\it Gauge Fields and Strings},  Harwood Academic Publishers, Chur (1087).
\bibitem{sim}For a review see: B. Simons, {\it Phase Transitions and Collective Phenomena}, Cambridge University Press, Cambridge (1997). 
\bibitem{klei}H. Kleinert, {\it Phys. Rev. Lett.} {\bf 58} (1987) 1915; F. David and E. Guitter, {\it Nucl. Phys.} {\bf B486} (1988) 332.
\bibitem{kle2}For a review see, e.g.: H. Kleinert, {\it Gauge Fields in Condensed Matter}, World Scientific, Singapore (1989). 

\end{references}
\end{document}